\documentclass[groupedaddress,twocolumn, aps, prl, lengthcheck, final]{revtex4-1}
\usepackage{amsmath,amssymb,graphicx}
\newcommand{\ket}[1]{\left |#1\right\rangle}
\newcommand{\down}{\ket{\downarrow}}
\newcommand{\up}{\ket{\uparrow}}
\begin{document}

\title{Extended coherence time on the clock transition of optically trapped Rubidium}
\author{G.~\surname{Kleine~B\"uning}$^{1}$}
\email{kleinebuening@iqo.uni-hannover.de}
\author{J.~Will$^{1}$}
\author{W.~Ertmer$^{1}$}
\author{E.~Rasel$^{1}$}
\author{J.~Arlt$^{2}$}
\author{C.~Klempt$^{1}$}
\affiliation{$^{1}$ Institut f\"ur Quantenoptik, Leibniz Universit\"at Hannover, Welfengarten 1, 30167 Hannover, Germany} 
\affiliation{$^{2}$ QUANTOP, Institut for Fysik og Astronomi, Aarhus Universitet, Ny Munkegade 120, 8000 Aarhus C, Denmark} 
\author{F.~\surname{Ramirez-Martinez}$^{3}$}
\author{F.~Pi\'echon$^{4}$}
\author{P.~Rosenbusch$^{3}$}
\affiliation{$^{3}$LNE-SYRTE, Observatoire de Paris, CNRS, UPMC, 61 av de l'Observatoire, 75014 Paris, France}
\affiliation{$^{4}$Laboratoire de Physique des Solides, CNRS, UMR 8502, Univ. Paris-Sud, 91405 Orsay, France}

\date{\today}

\begin{abstract}
Optically trapped ensembles are of crucial importance for frequency measurements and quantum memories, but generally suffer from strong dephasing due to inhomogeneous density and light shifts. We demonstrate a drastic increase of the coherence time to $21$~s on the magnetic field insensitive clock transition of $^{87}$Rb by applying the recently discovered spin self-rephasing~\cite{deutsch2010}. This result confirms the general nature of this new mechanism and thus shows its applicability in atom clocks and quantum memories. A systematic investigation of all relevant frequency shifts and noise contributions yields a stability of $2.4\times 10^{-11}\tau^{-1/2}$, where $\tau$ is the integration time in seconds. Based on a set of technical improvements, the presented frequency standard is predicted to rival the stability of microwave fountain clocks in a potentially much more compact setup.
\end{abstract}

\maketitle

Atomic clocks provide unprecedented precision in the determination of time. Most prominently, the SI second is experimentally determined by measuring the $^{133}$Cs hyperfine transition with microwave atomic fountain clocks~\cite{Wynands2005}. The ongoing quest for the best definition of time is currently led by optical frequency standards~\cite{Chou2010,Ludlow2008}. Moreover, applications ranging from the determination of fundamental constants to navigation and communication technology demand more compact frequency standards. In contrast to microwave fountain clocks, optically trapped atomic ensembles~\cite{[{Note recent progress in optically trapped Bose-Einstein condensates as reported in }][{}]Altin2010} promise high stability in compact setups due to long interrogation times. However, they typically suffer from a short coherence time.

In addition to frequency measurements, trapped ensembles serve as efficient quantum memories for long-distance quantum communication~\cite{Hammerer2010}. For quantum memories, light is coupled to spin waves on the hyperfine transition of alkali atoms. In a series of experiments, deterministic photons were produced~\cite{Matsukevich2006}, entanglement between photons and spin waves was created~\cite{Dudin2009,Riedmatten2006}, and even prototype networks for entanglement distribution were demonstrated~\cite{Chou2007,Papp2009}.  By employing traps instead of ballistic ensembles, the memory lifetime was already extended from microseconds~\cite{Matsukevich2006} to milliseconds~\cite{Zhao2009a,Zhao2009}. Longer coherence times would also be beneficial in this field.

The coherence time is limited by dephasing due to trap-induced frequency shifts which are necessarily inhomogeneous. There have been several attempts to overcome the influence of these shifts~\cite{Derevianko2010}. Optical lattice clocks employ a dipole trap at a magic wavelength to suppress the differential light shift (DLS)~\cite{Katori2003}. Magic wavelengths for microwave transitions have been predicted for Al and Ga \cite{Beloy2009}. However, the microwave transitions in alkali atoms such as Rb and Cs do not exhibit magic wavelengths~\cite{Rosenbusch2009} unless traps of particular polarization combined with a specific magnetic field are employed~\cite{Derevianko2010a,Chicireanu2011}. There have been approaches to increase the coherence time by collisional spectral narrowing~\cite{davidson.105.093001} or artificial rephasing through a sequence of microwave pulses~\cite{davidson.105.053201}.  A further approach is the use of a magnetic confinement for which the inhomogeneity of the Zeeman shift is canceled by the density shift~\cite{Lewandowski2002}. In such a trap, it was shown that the exchange interaction can be used to induce spin self-rephasing (SSR)~\cite{deutsch2010}. It is therefore an important question whether this effect may also be used for the magnetically insensitive clock states.

In this Letter, we employ spin self-rephasing to extend the coherence time of optically trapped atoms to $21$ seconds. We show that both the inhomogeneity of the DLS and the inhomogeneity of the density shift can be overcome by SSR. Within our investigation, we perform Ramsey spectroscopy using the magnetically insensitive clock states of the $^{87}$Rb hyperfine transition, a secondary representation of the second~\cite{Rbsecondary}. Thus, we evaluate the applicability of SSR for atomic microwave frequency standards and demonstrate a stability of $2.4\times10^{-11}\tau^{-1/2}$. For a purpose-built apparatus with standard technical improvements, we predict a 300\hbox{-}fold stability enhancement, primarily limited by temperature fluctuations. Thereby our technique enters the stability range of most atomic fountain clocks~\cite{Wynands2005} in a potentially more compact setup. Our findings prove the predicted fundamental nature of SSR which is independent of the type of trap and the magnetic quantum number. Thus, its applicability extends to optical lattice clocks as well as quantum information with atomic ensembles. 

\begin{figure*}
\includegraphics[width=\textwidth]{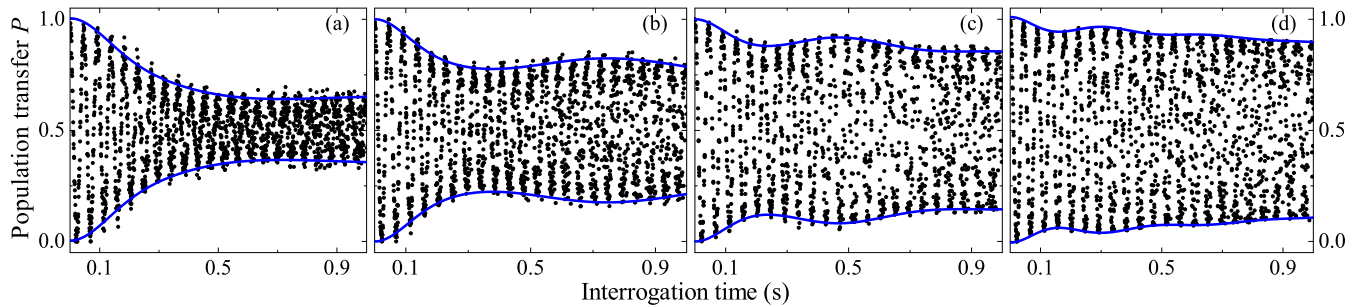}
\caption{Population transfer $P\!=\!N_\uparrow/(N_\uparrow+N_\downarrow)$ after a Ramsey sequence as a function of the interrogation time. The mean density increasing from (a) to (d) is $\{0.18, 0.36, 0.65, 1.38\}\times 10^{12}$~cm$^{-3}$. Both the increase of the coherence time and the characteristic contrast revivals due to spin self-rephasing are clearly visible.  Further revivals at longer interrogation times are damped out due to a mixing of trap states. Both conditions for spin self-rephasing are fulfilled: $\omega_\text{ex}/\Delta_0$ increases with density from $4.3$ to $23$, giving rise to the observed onset of rephasing, while $\omega_\text{ex}/(\pi \gamma_\text{c})\!=\!4.8$ does not depend on density. The solid lines indicate fits according to the numeric model (see text), closely reproducing the characteristic behavior of the contrast.}
\label{FigRevivals}
\end{figure*}

The SSR inhibits dephasing and induces characteristic contrast revivals, mediated by the exchange interaction. This effect can be understood as follows. Each atom in a superposition of the two clock states, can be represented by an effective spin~$1/2$. The exchange interaction of two atoms can be described within the singlet/triplet basis~\cite{gibble2009,gibble2010}. The atom pair, initially in the triplet state, dephases by Rabi flopping towards the singlet state. For bosons (fermions), the interaction energy shifts solely the triplet (singlet) state due to the symmetry of the spatial wave function. The shift tunes the states out of resonance and thereby suppresses the Rabi flopping and hence the dephasing. The effect can also be visualized in the Bloch sphere representation, where the exchange interaction corresponds to a rotation of two individual spins around their sum~\cite{Fuchs2002}. In the following, we apply the quantitative model developed in Ref.~\cite{deutsch2010}. In this approach, the atoms are divided into classes of constant energy and a mean spin for each class is considered. 

Two conditions have to be met for SSR to occur: (i) the inhomogeneity of the transition frequency, i.e. the dephasing rate, must be smaller than the exchange rate, $\Delta_0<\omega_\text{ex}$; (ii) the rate of velocity changing collisions must be smaller than the exchange rate so that rephasing establishes before atoms change energy class, $\gamma_\text{c}<\omega_\text{ex}/\pi$. These three parameters can be predicted for a thermal gas in a harmonic potential. We use the definition of $\Delta_0$, $\gamma_\text{c}$ and $\omega_\text{ex}$ from~\cite{deutsch2010}. In our case, the dephasing rate $\Delta_0$ is composed of two inhomogeneous frequency shifts, the differential light shift and the density shift in the mean-field approximation:   
$\Delta_0\!=\!k_\text{B} T/2 \hbar \times \delta\alpha/\alpha-\gamma\bar{n}/4$. Here, $\delta \alpha$ and $\alpha$ denote the differential and the total light shift per intensity, $T$ is the ensemble temperature, $\bar n$ the mean density and $\gamma\!=\!\frac{4\pi\hbar}{m}(a_{\uparrow\uparrow}-a_{\downarrow\downarrow})$, where $m$ is the mass and $a_{\uparrow\uparrow}$, $a_{\downarrow\downarrow}$ are the scattering lengths between atoms in the same state.

In a first set of experiments, we explore the onset of SSR in an optically trapped ensemble. Our experimental apparatus has been described in detail in Ref.~\cite{kleinebuening2010}. An optical dipole trap at $\lambda\!=\!1064$~nm is provided by two intersecting laser beams with beam waists of 60 and 75~$\mu$m, yielding a mean trap frequency of $2 \pi \times 70$~Hz. A cloud of $^{87}$Rb-atoms is prepared in the state $\down\!=\!\ket{F\!=\!1, m_F\!=\!0}$ at a temperature of $400(20)$~nK with a variable mean density $\bar n$ from $0.18$ to $1.38 \times 10^{12}$~cm$^{-3}$. The atoms are held in the trap for a constant hold time of $1.06$~s, during which Ramsey spectroscopy of the clock transition to $\up\!=\!\ket{F\!=\!2, m_F\!=\!0}$ is performed. The Ramsey sequence with a variable interrogation time of $0-1$~s is placed at the end within this hold time. The coupling is realized by two microwave $\pi/2$-pulses of $100~\mu$s duration, generated by a low phase noise synthesizer~\cite{Ramirez-Martinez2010} locked to a hydrogen maser. After the Ramsey sequence, the trap is switched off to allow for ballistic expansion and Stern-Gerlach separation of the two states. The number of atoms in both states, $N_\uparrow$ and $N_\downarrow$, is detected by simultaneous absorption imaging, calibrated according to Ref.~\cite{Reinaudi2007}.

Figure~\ref{FigRevivals} shows the population transfer $P\!=\!N_\uparrow/(N_\uparrow+N_\downarrow)$ as a function of the interrogation time for four different densities. Each subplot shows clear sinusoidal Ramsey fringes with a period of $\approx\! 45$~ms, in agreement with a microwave detuning of $15$~Hz and residual shifts of $-7$~Hz (see below). Throughout this work, the microwave detuning is referenced to the hyperfine frequency at zero field~\cite{Guena2010}. For the lowest density, the contrast of the Ramsey fringes drops within the first $500$~ms due to dephasing. However, the contrast does not decay completely within $1$~s, as expected for pure dephasing, indicating the presence of SSR. At higher mean density, a revival of the contrast emerges and a considerably longer coherence time is observed. As the density is increased further, the contrast revivals appear at earlier times, since the higher exchange rate $\omega_\text{ex}$ ensures faster rephasing. The contrast does not recover fully at later times, because the rate of velocity changing collisions $\gamma_\text{c}$ abates rephasing by a mixing of trap states. 

The measured contrast is well reproduced by the numerical model of Ref.~\cite{deutsch2010}, where $\Delta_0$, $\omega_\text{ex}$ and $\gamma_\text{c}$ are used as free fitting parameters. They agree within 25\% with their prediction if the same correction factors of $1.6$ for $\Delta_0$ and $0.6$ for $\omega_\text{ex}$ are included. For the case of highest density ($1.38 \times 10^{12}$~cm$^{-3}$) the agreement is within 50\%, which indicates failure of the infinite-range approximation. Note that the determination of $\Delta_0$ necessitates a good knowledge of the DLS and the density shift (see below). The simple theoretical model provides an excellent understanding of the measurements and allows for an optimization of the relevant experimental parameters towards a drastic increase of the coherence time.

\begin{figure}
\includegraphics[width= \columnwidth]{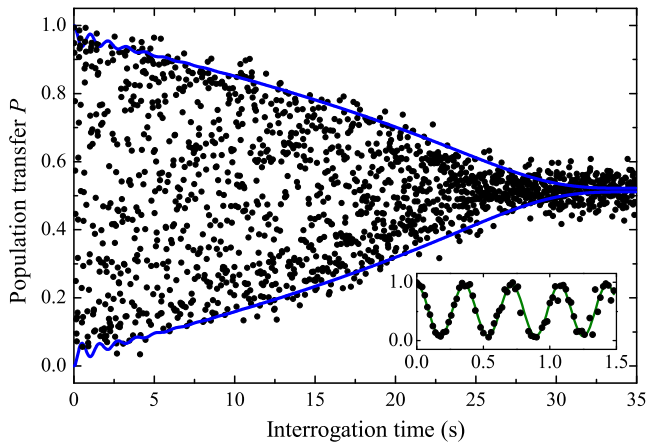}
\caption{Population transfer $P$ in a Ramsey sequence optimized for long coherence time (see text). The inset shows clear fringes for the first 1.5~s. However, they wash out due to frequency noise after some seconds. Nonetheless the contrast remains well visible and a $1/e$ coherence time of 21~s is obtained, which constitutes a dramatic increase over previous experiments. The solid line represents a solution of the numeric model.}
\label{FigLongTimes} 
\end{figure}

Improvement of the coherence time requires simultaneous optimization of conditions (i) and (ii). While $\omega_\text{ex}/\Delta_0$ can easily be increased with density, $\omega_\text{ex}/(\pi \gamma_\text{c})$ can only be increased by lowering the temperature. The mean trap frequency is therefore reduced to $2 \pi \times 51$~Hz while cooling the ensemble to a temperature of $100(20)$~nK at a mean density of $0.3 \times 10^{12}$cm$^{-3}$. In such an optimized configuration, $\omega_\text{ex}/(\pi \gamma_\text{c})$ is increased to $9.6$ while preserving $\Delta_0/\omega_\text{ex}$ at $20.8$. Under these conditions, the coherence time is investigated within an extended hold time of $36.2$~s. Figure~\ref{FigLongTimes} shows the population transfer $P$ for a variable interrogation time with a microwave detuning of $0$~Hz. Although the fringes are quickly washed out due to the instability of the absolute frequency, the contrast persists over many seconds. The $1/e$ coherence time is 21~s and complete loss of coherence occurs at 26~s. A reduction of the contrast due to asymmetric atom loss can be neglected since we measure a 38~s differential lifetime of $\up$ and $\down$. The measured contrast is well reproduced by the numerical model without parameter fitting. Note that the model reflects the non-exponential shape of the contrast reduction.

\begin{figure}
\includegraphics[width=\columnwidth]{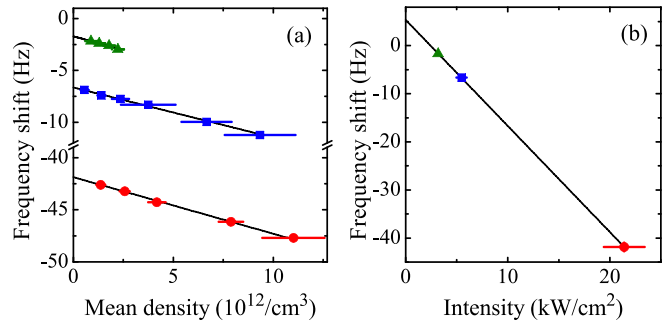}
\caption{Measurement of the relevant frequency shifts compared to the zero field transition. Three trap configurations with mean trap frequencies of $2\pi \times 51$~Hz (green triangles), $2\pi \times 70$~Hz (blue squares), and $2\pi \times 112$~Hz (red circles) were investigated. (a) Frequency shift as a function of the mean density (statistical error bars). The extrapolation to zero density yields a residual shift, plotted in (b) as a function of the ensemble averaged light intensity for each trap (estimated uncertainties). The slope of a linear fit determines the DLS.}
\label{FigShifts}
\end{figure}

The presented analysis of spin self-rephasing requires precise knowledge of both the mean field density shift and the DLS. To determine these values, we perform measurements of the absolute frequency for various atomic densities in three different trap configurations with mean trap frequencies of  $2 \pi \times $\{51,70, and 112\}~Hz. The measured frequency shifts are presented in Fig.~\ref{FigShifts}(a). All three configurations show a clear linear dependence on the mean density. The fitted slopes lie within 10\% of $\gamma/2\pi\!=\! -0.52$~Hz$/(10^{12}$cm$^{-3})$ and agree well with the predicted value of $-0.48$~Hz$/(10^{12}$cm$^{-3})$~\footnote{We use a$_{\uparrow\uparrow}\!=\!94.55~a_\text{B}$, a$_{\downarrow\downarrow}\!=\!100.76~a_\text{B}$ from a coupled channel analysis by E. Tiemann.}. The DLS can be obtained from an extrapolation of this frequency shift to zero density. Figure~\ref{FigShifts}(b) shows this value as a function of the ensemble averaged light intensity. We find a linear dependence on the intensity with  $\delta\alpha/2 \pi\!=\!-2.2(3)$~Hz/(kW~cm$^{-2}$), which constitutes the first direct measurement of the $^{87}$Rb clock shift at $1064$~nm. This value confirms a first principle calculation~\footnote{A.~Derevianko, Priv. Comm. (see also \cite{Derevianko2010a})} yielding $-2.4$~Hz/(kW~cm$^{-2}$). The extrapolation to zero intensity results in a residual shift of $5.2(6)$~Hz. This value is in excellent agreement with the quadratic Zeeman shift of $5.2(1)$~Hz at the magnetic offset field of $95(1)$~mG which is determined from the $\Delta m_F\!=\!1$ transition. The presented extrapolation procedure will allow for an absolute measurement of the $^{87}$Rb clock frequency in an optical trap.

\begin{figure}
\includegraphics[width=\columnwidth]{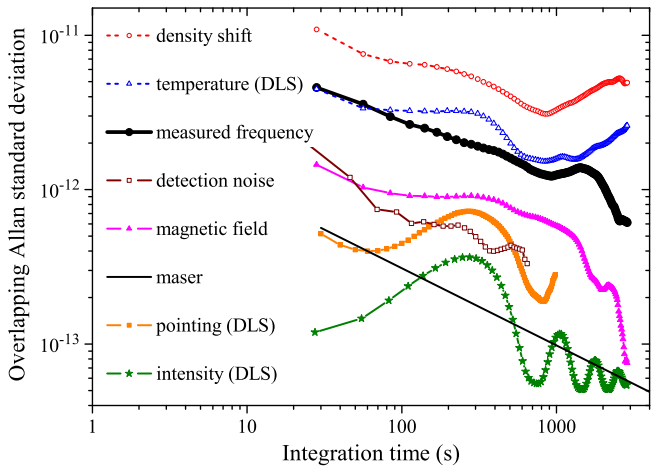}
\caption{Stability analysis based on the overlapping Allan standard deviation. The frequency measurement (full black dots) yields a stability of $4.6\times 10^{-12}$ at one shot, corresponding to $2.4\times 10^{-11}$ at one second. All relevant noise sources are measured independently (see text) and are shown here as frequency noise. The periodic modulation in some of the curves reflects the air-conditioning cycle of 10~minutes.}
\label{FigAllanDeviation}
\end{figure}

Finally, we present the frequency stability of our measurements  and analyze its principal noise contributions. Thereby, we can infer the potential benefit of long coherence times for a microwave frequency standard using an optically trapped ensemble in a purpose-built setup. To determine the noise contributions, the frequency is measured by Ramsey spectroscopy in a similar configuration to Fig.~\ref{FigRevivals}(d). The interrogation time was set to $350$~ms coinciding with the first contrast revival. The detuning of $-8.4$~Hz was chosen such that the mean population transfer is $P\!\approx\! 0.5$ on the central Ramsey fringe. The population transfer $P$ was recorded in 215 consecutive measurements at a cycle time of $28.1$~s. The number of measurements was limited to $\approx\! 2$ hours at night during standstill of the near-by tram line. Figure~\ref{FigAllanDeviation} shows the overlapping Allan standard deviation~\cite{Howe1981} for the frequencies extracted from the population transfer. The data follows a near-white frequency noise behavior of $2.4\times 10^{-11}\tau^{-1/2}$, equivalent to $4.6\times 10^{-12}$ at one shot.

For our parameters, the quantum projection noise limit is $8\times 10^{-13}\tau^{-1/2}$, showing that our measurements are limited by technical noise. Seven relevant noise sources are measured independently and their frequency equivalents are plotted in Figure~\ref{FigAllanDeviation}: The maser is not limiting our measurements as its typical frequency stability shows. The detection noise is $1\%$ at one shot. Both of these effects decrease with a higher duty cycle and may be suppressed substantially by the prolonged interrogation time due to SSR. The DLS fluctuates if intensity or pointing of the trapping beams change, which were measured as $6\times10^{-5}$ and $0.3~\mu$m at one shot, respectively. The magnetic field contribution, determined to be $0.1$~mG at one shot with an external Hall probe, is significant due to the lack of shielding. Figure~\ref{FigAllanDeviation} also shows that the mean field density noise and the DLS noise due to temperature fluctuations are overestimated due to their imprecise deduction from absorption images.

In summary, the effect of SSR was investigated in an optically trapped ensemble of $^{87}$Rb. By engineering the spin self-rephasing, we have demonstrated a coherence time of $21$~s -- the longest ensemble coherence time reported so far. The shifts and noise contributions to a frequency measurement were studied systematically and a stability of $2.4\times 10^{-11}\tau^{-1/2}$ was reached. For an optimized setup including magnetic field shielding, shallower trapping, a 10\hbox{-}fold increase in temperature determination and an increase in duty cycle exploiting the full coherence time, we predict an achievable stability of $8 \times 10^{-14} \tau^{-1/2}$. This rivals the stability range of most atomic fountain clocks~\cite{Wynands2005} in a potentially much more compact setup. The proven generality of the giant coherence time due to SSR makes it a promising candidate for application in optical lattice clocks and ensemble-based quantum information.

\begin{acknowledgments}
We acknowledge support from the Centre for Quantum Engineering and Space-Time Research (QUEST), the Institut Francilien
de Recherche sur les Atomes Froids (IFRAF) and the Danish National Research Foundation Center for Quantum Optics (QUANTOP).
\end{acknowledgments}

\bibliography{kleinebuening}

\begin{thebibliography}{10}%
\makeatletter
\providecommand \@ifxundefined [1]{%
 \ifx #1\undefined \expandafter \@firstoftwo
 \else \expandafter \@secondoftwo
\fi
}%
\providecommand \@ifnum [1]{%
 \ifnum #1\expandafter \@firstoftwo
 \else \expandafter \@secondoftwo
\fi
}%
\providecommand \enquote [1]{``#1''}%
\providecommand \bibnamefont  [1]{#1}%
\providecommand \bibfnamefont [1]{#1}%
\providecommand \citenamefont [1]{#1}%
\providecommand\href[0]{\@sanitize\@href}%
\providecommand\@href[1]{\endgroup\@@startlink{#1}\endgroup\@@href}%
\providecommand\@@href[1]{#1\@@endlink}%
\providecommand \@sanitize [0]{\begingroup\catcode`\&12\catcode`\#12\relax}%
\@ifxundefined \pdfoutput {\@firstoftwo}{%
 \@ifnum{\z@=\pdfoutput}{\@firstoftwo}{\@secondoftwo}%
}{%
 \providecommand\@@startlink[1]{\leavevmode\special{html:<a href="#1">}}%
 \providecommand\@@endlink[0]{\special{html:</a>}}%
}{%
 \providecommand\@@startlink[1]{%
  \leavevmode
  \pdfstartlink
   attr{/Border[0 0 1 ]/H/I/C[0 1 1]}%
   user{/Subtype/Link/A<</Type/Action/S/URI/URI(#1)>>}%
  \relax
 }%
 \providecommand\@@endlink[0]{\pdfendlink}%
}%
\providecommand \url  [0]{\begingroup\@sanitize \@url }%
\providecommand \@url [1]{\endgroup\@href {#1}{\urlprefix}}%
\providecommand \urlprefix [0]{URL }%
\providecommand \Eprint[0]{\href }%
\@ifxundefined \urlstyle {%
  \providecommand \doi [1]{doi:\discretionary{}{}{}#1}%
}{%
  \providecommand \doi [0]{doi:\discretionary{}{}{}\begingroup
  \urlstyle{rm}\Url }%
}%
\providecommand \doibase [0]{http://dx.doi.org/}%
\providecommand \Doi[1]{\href{\doibase#1}}%
\providecommand \bibAnnote [3]{%
  \BibitemShut{#1}%
  \begin{quotation}\noindent
    \textsc{Key:}\ #2\\\textsc{Annotation:}\ #3%
  \end{quotation}%
}%
\providecommand \bibAnnoteFile [2]{%
  \IfFileExists{#2}{\bibAnnote {#1} {#2} {\input{#2}}}{}%
}%
\providecommand \typeout [0]{\immediate \write \m@ne }%
\providecommand \selectlanguage [0]{\@gobble}%
\providecommand \bibinfo [0]{\@secondoftwo}%
\providecommand \bibfield [0]{\@secondoftwo}%
\providecommand \translation [1]{[#1]}%
\providecommand \BibitemOpen[0]{}%
\providecommand \bibitemStop [0]{}%
\providecommand \bibitemNoStop [0]{.\EOS\space}%
\providecommand \EOS [0]{\spacefactor3000\relax}%
\providecommand \BibitemShut [1]{\csname bibitem#1\endcsname}%
\bibitem{deutsch2010}%
  \BibitemOpen
  \bibfield{author}{%
  \bibinfo {author} {\bibfnamefont{C.}~\bibnamefont{Deutsch}}, \bibinfo
  {author} {\bibfnamefont{F.}~\bibnamefont{Ramirez-Martinez}}, \bibinfo
  {author} {\bibfnamefont{C.}~\bibnamefont{Lacro\^ute}}, \bibinfo {author}
  {\bibfnamefont{F.}~\bibnamefont{Reinhard}}, \bibinfo {author}
  {\bibfnamefont{T.}~\bibnamefont{Schneider}}, \bibinfo {author}
  {\bibfnamefont{J.~N.}\ \bibnamefont{Fuchs}}, \bibinfo {author}
  {\bibfnamefont{F.}~\bibnamefont{Pi\'echon}}, \bibinfo {author}
  {\bibfnamefont{F.}~\bibnamefont{Lalo\"e}}, \bibinfo {author}
  {\bibfnamefont{J.}~\bibnamefont{Reichel}},\ and\ \bibinfo {author}
  {\bibfnamefont{P.}~\bibnamefont{Rosenbusch}},\ }%
  \bibfield{journal}{%
  \Doi{10.1103/PhysRevLett.105.020401}{\bibinfo {journal} {Phys. Rev. Lett.}}\
  }%
  \textbf{\bibinfo {volume} {105}},\ \bibinfo {pages} {020401} (\bibinfo {year}
  {2010})%
  \bibAnnoteFile{NoStop}{deutsch2010}%
\bibitem{Wynands2005}%
  \BibitemOpen
  \bibfield{author}{%
  \bibinfo {author} {\bibfnamefont{R.}~\bibnamefont{Wynands}}\ and\ \bibinfo
  {author} {\bibfnamefont{S.}~\bibnamefont{Weyers}},\ }%
  \bibfield{journal}{%
  \bibinfo {journal} {Metrologia}\ }%
  \textbf{\bibinfo {volume} {42}},\ \bibinfo {pages} {S64} (\bibinfo {year}
  {2005})%
  \bibAnnoteFile{NoStop}{Wynands2005}%
\bibitem{Chou2010}%
  \BibitemOpen
  \bibfield{author}{%
  \bibinfo {author} {\bibfnamefont{C.~W.}\ \bibnamefont{Chou}}, \bibinfo
  {author} {\bibfnamefont{D.~B.}\ \bibnamefont{Hume}}, \bibinfo {author}
  {\bibfnamefont{J.~C.~J.}\ \bibnamefont{Koelemeij}}, \bibinfo {author}
  {\bibfnamefont{D.~J.}\ \bibnamefont{Wineland}},\ and\ \bibinfo {author}
  {\bibfnamefont{T.}~\bibnamefont{Rosenband}},\ }%
  \bibfield{journal}{%
  \Doi{10.1103/PhysRevLett.104.070802}{\bibinfo {journal} {Phys. Rev. Lett.}}\
  }%
  \textbf{\bibinfo {volume} {104}},\ \bibinfo {pages} {070802} (\bibinfo {year}
  {2010})%
  \bibAnnoteFile{NoStop}{Chou2010}%
\bibitem{Ludlow2008}%
  \BibitemOpen
  \bibfield{author}{%
  \bibinfo {author} {\bibfnamefont{A.~D.}\ \bibnamefont{Ludlow}}, \bibinfo
  {author} {\bibfnamefont{T.}~\bibnamefont{Zelevinsky}}, \bibinfo {author}
  {\bibfnamefont{G.~K.}\ \bibnamefont{Campbell}}, \bibinfo {author}
  {\bibfnamefont{S.}~\bibnamefont{Blatt}}, \bibinfo {author}
  {\bibfnamefont{M.~M.}\ \bibnamefont{Boyd}}, \bibinfo {author}
  {\bibfnamefont{M.~H.~G.}\ \bibnamefont{de~Miranda}}, \bibinfo {author}
  {\bibfnamefont{M.~J.}\ \bibnamefont{Martin}}, \bibinfo {author}
  {\bibfnamefont{J.~W.}\ \bibnamefont{Thomsen}}, \bibinfo {author}
  {\bibfnamefont{S.~M.}\ \bibnamefont{Foreman}}, \bibinfo {author}
  {\bibfnamefont{J.}~\bibnamefont{Ye}}, \bibinfo {author}
  {\bibfnamefont{T.~M.}\ \bibnamefont{Fortier}}, \bibinfo {author}
  {\bibfnamefont{J.~E.}\ \bibnamefont{Stalnaker}}, \bibinfo {author}
  {\bibfnamefont{S.~A.}\ \bibnamefont{Diddams}}, \bibinfo {author}
  {\bibfnamefont{Y.}~\bibnamefont{Le~Coq}}, \bibinfo {author}
  {\bibfnamefont{Z.~W.}\ \bibnamefont{Barber}}, \bibinfo {author}
  {\bibfnamefont{N.}~\bibnamefont{Poli}}, \bibinfo {author}
  {\bibfnamefont{N.~D.}\ \bibnamefont{Lemke}}, \bibinfo {author}
  {\bibfnamefont{K.~M.}\ \bibnamefont{Beck}},\ and\ \bibinfo {author}
  {\bibfnamefont{C.~W.}\ \bibnamefont{Oates}},\ }%
  \bibfield{journal}{%
  \Doi{10.1126/science.1153341}{\bibinfo {journal} {Science}}\ }%
  \textbf{\bibinfo {volume} {319}},\ \bibinfo {pages} {1805} (\bibinfo {year}
  {2008})%
  \bibAnnoteFile{NoStop}{Ludlow2008}%
\bibitem{Altin2010}%
  \BibitemOpen
  \bibfield{author}{%
  \bibinfo {author} {\bibfnamefont{P.~A.}\ \bibnamefont{{Altin}}}, \bibinfo
  {author} {\bibfnamefont{G.}~\bibnamefont{{McDonald}}}, \bibinfo {author}
  {\bibfnamefont{D.}~\bibnamefont{{D{\"o}ring}}}, \bibinfo {author}
  {\bibfnamefont{J.~E.}\ \bibnamefont{{Debs}}}, \bibinfo {author}
  {\bibfnamefont{T.}~\bibnamefont{{Barter}}}, \bibinfo {author}
  {\bibfnamefont{N.~P.}\ \bibnamefont{{Robins}}}, \bibinfo {author}
  {\bibfnamefont{J.~D.}\ \bibnamefont{{Close}}}, \bibinfo {author}
  {\bibfnamefont{S.~A.}\ \bibnamefont{{Haine}}}, \bibinfo {author}
  {\bibfnamefont{T.~M.}\ \bibnamefont{{Hanna}}},\ and\ \bibinfo {author}
  {\bibfnamefont{R.~P.}\ \bibnamefont{{Anderson}}},\ }%
  \enquote{\bibinfo {title} {{Optically trapped atom interferometry using the
  clock transition of large $^{87}$Rb Bose-Einstein condensates}},}\  (\bibinfo
  {year} {2010}),\ \Eprint{http://arxiv.org/abs/1011.4713}{arXiv:1011.4713}%
  \bibAnnoteFile{NoStop}{Altin2010}%
\bibitem{Hammerer2010}%
  \BibitemOpen
  \bibfield{author}{%
  \bibinfo {author} {\bibfnamefont{K.}~\bibnamefont{Hammerer}}, \bibinfo
  {author} {\bibfnamefont{A.~S.}\ \bibnamefont{S\o{}rensen}},\ and\ \bibinfo
  {author} {\bibfnamefont{E.~S.}\ \bibnamefont{Polzik}},\ }%
  \bibfield{journal}{%
  \Doi{10.1103/RevModPhys.82.1041}{\bibinfo {journal} {Rev. Mod. Phys.}}\ }%
  \textbf{\bibinfo {volume} {82}},\ \bibinfo {pages} {1041} (\bibinfo {year}
  {2010})%
  \bibAnnoteFile{NoStop}{Hammerer2010}%
\bibitem{Matsukevich2006}%
  \BibitemOpen
  \bibfield{author}{%
  \bibinfo {author} {\bibfnamefont{D.~N.}\ \bibnamefont{Matsukevich}}, \bibinfo
  {author} {\bibfnamefont{T.}~\bibnamefont{Chaneli\`ere}}, \bibinfo {author}
  {\bibfnamefont{S.~D.}\ \bibnamefont{Jenkins}}, \bibinfo {author}
  {\bibfnamefont{S.-Y.}\ \bibnamefont{Lan}}, \bibinfo {author}
  {\bibfnamefont{T.~A.~B.}\ \bibnamefont{Kennedy}},\ and\ \bibinfo {author}
  {\bibfnamefont{A.}~\bibnamefont{Kuzmich}},\ }%
  \bibfield{journal}{%
  \Doi{10.1103/PhysRevLett.97.013601}{\bibinfo {journal} {Phys. Rev. Lett.}}\
  }%
  \textbf{\bibinfo {volume} {97}},\ \bibinfo {pages} {013601} (\bibinfo {year}
  {2006})%
  \bibAnnoteFile{NoStop}{Matsukevich2006}%
\bibitem{Dudin2009}%
  \BibitemOpen
  \bibfield{author}{%
  \bibinfo {author} {\bibfnamefont{Y.~O.}\ \bibnamefont{Dudin}}, \bibinfo
  {author} {\bibfnamefont{S.~D.}\ \bibnamefont{Jenkins}}, \bibinfo {author}
  {\bibfnamefont{R.}~\bibnamefont{Zhao}}, \bibinfo {author}
  {\bibfnamefont{D.~N.}\ \bibnamefont{Matsukevich}}, \bibinfo {author}
  {\bibfnamefont{A.}~\bibnamefont{Kuzmich}},\ and\ \bibinfo {author}
  {\bibfnamefont{T.~A.~B.}\ \bibnamefont{Kennedy}},\ }%
  \bibfield{journal}{%
  \Doi{10.1103/PhysRevLett.103.020505}{\bibinfo {journal} {Phys. Rev. Lett.}}\
  }%
  \textbf{\bibinfo {volume} {103}},\ \bibinfo {pages} {020505} (\bibinfo {year}
  {2009})%
  \bibAnnoteFile{NoStop}{Dudin2009}%
\bibitem{Riedmatten2006}%
  \BibitemOpen
  \bibfield{author}{%
  \bibinfo {author} {\bibfnamefont{H.}~\bibnamefont{de~Riedmatten}}, \bibinfo
  {author} {\bibfnamefont{J.}~\bibnamefont{Laurat}}, \bibinfo {author}
  {\bibfnamefont{C.~W.}\ \bibnamefont{Chou}}, \bibinfo {author}
  {\bibfnamefont{E.~W.}\ \bibnamefont{Schomburg}}, \bibinfo {author}
  {\bibfnamefont{D.}~\bibnamefont{Felinto}},\ and\ \bibinfo {author}
  {\bibfnamefont{H.~J.}\ \bibnamefont{Kimble}},\ }%
  \bibfield{journal}{%
  \Doi{10.1103/PhysRevLett.97.113603}{\bibinfo {journal} {Phys. Rev. Lett.}}\
  }%
  \textbf{\bibinfo {volume} {97}},\ \bibinfo {pages} {113603} (\bibinfo {year}
  {2006})%
  \bibAnnoteFile{NoStop}{Riedmatten2006}%
\bibitem{Chou2007}%
  \BibitemOpen
  \bibfield{author}{%
  \bibinfo {author} {\bibfnamefont{C.-W.}\ \bibnamefont{Chou}}, \bibinfo
  {author} {\bibfnamefont{J.}~\bibnamefont{Laurat}}, \bibinfo {author}
  {\bibfnamefont{H.}~\bibnamefont{Deng}}, \bibinfo {author}
  {\bibfnamefont{K.~S.}\ \bibnamefont{Choi}}, \bibinfo {author}
  {\bibfnamefont{H.}~\bibnamefont{de~Riedmatten}}, \bibinfo {author}
  {\bibfnamefont{D.}~\bibnamefont{Felinto}},\ and\ \bibinfo {author}
  {\bibfnamefont{H.~J.}\ \bibnamefont{Kimble}},\ }%
  \bibfield{journal}{%
  \Doi{10.1126/science.1140300}{\bibinfo {journal} {Science}}\ }%
  \textbf{\bibinfo {volume} {316}},\ \bibinfo {pages} {1316} (\bibinfo {year}
  {2007})%
  \bibAnnoteFile{NoStop}{Chou2007}%
\bibitem{Papp2009}%
  \BibitemOpen
  \bibfield{author}{%
  \bibinfo {author} {\bibfnamefont{S.~B.}\ \bibnamefont{Papp}}, \bibinfo
  {author} {\bibfnamefont{K.~S.}\ \bibnamefont{Choi}}, \bibinfo {author}
  {\bibfnamefont{H.}~\bibnamefont{Deng}}, \bibinfo {author}
  {\bibfnamefont{P.}~\bibnamefont{Lougovski}}, \bibinfo {author}
  {\bibfnamefont{S.~J.}\ \bibnamefont{van Enk}},\ and\ \bibinfo {author}
  {\bibfnamefont{H.~J.}\ \bibnamefont{Kimble}},\ }%
  \bibfield{journal}{%
  \bibinfo {journal} {Science}\ }%
  \textbf{\bibinfo {volume} {324}},\ \bibinfo {pages} {764} (\bibinfo {year}
  {2009})%
  \bibAnnoteFile{NoStop}{Papp2009}%
\bibitem{Zhao2009a}%
  \BibitemOpen
  \bibfield{author}{%
  \bibinfo {author} {\bibfnamefont{B.}~\bibnamefont{Zhao}}, \bibinfo {author}
  {\bibfnamefont{Y.-A.}\ \bibnamefont{Chen}}, \bibinfo {author}
  {\bibfnamefont{X.-H.}\ \bibnamefont{Bao}}, \bibinfo {author}
  {\bibfnamefont{T.}~\bibnamefont{Strassel}}, \bibinfo {author}
  {\bibfnamefont{C.-S.}\ \bibnamefont{Chuu}}, \bibinfo {author}
  {\bibfnamefont{X.-M.}\ \bibnamefont{Jin}}, \bibinfo {author}
  {\bibfnamefont{J.}~\bibnamefont{Schmiedmayer}}, \bibinfo {author}
  {\bibfnamefont{Z.-S.}\ \bibnamefont{Yuan}}, \bibinfo {author}
  {\bibfnamefont{S.}~\bibnamefont{Chen}},\ and\ \bibinfo {author}
  {\bibfnamefont{J.-W.}\ \bibnamefont{Pan}},\ }%
  \bibfield{journal}{%
  \Doi{10.1038/nphys1153}{\bibinfo {journal} {Nat Phys}}\ }%
  \textbf{\bibinfo {volume} {5}},\ \bibinfo {pages} {95} (\bibinfo {year}
  {2009})%
  \bibAnnoteFile{NoStop}{Zhao2009a}%
\bibitem{Zhao2009}%
  \BibitemOpen
  \bibfield{author}{%
  \bibinfo {author} {\bibfnamefont{R.}~\bibnamefont{Zhao}}, \bibinfo {author}
  {\bibfnamefont{Y.}~\bibnamefont{Dudin}}, \bibinfo {author}
  {\bibfnamefont{S.}~\bibnamefont{Jenkins}}, \bibinfo {author}
  {\bibfnamefont{C.}~\bibnamefont{Campbell}}, \bibinfo {author}
  {\bibfnamefont{D.}~\bibnamefont{Matsukevich}}, \bibinfo {author}
  {\bibfnamefont{T.}~\bibnamefont{Kennedy}},\ and\ \bibinfo {author}
  {\bibfnamefont{A.}~\bibnamefont{Kuzmich}},\ }%
  \bibfield{journal}{%
  \Doi{10.1038/nphys1152}{\bibinfo {journal} {Nat Phys}}\ }%
  \textbf{\bibinfo {volume} {5}},\ \bibinfo {pages} {100} (\bibinfo {year}
  {2009})%
  \bibAnnoteFile{NoStop}{Zhao2009}%
\bibitem{Derevianko2010}%
  \BibitemOpen
  \bibfield{author}{%
  \bibinfo {author} {\bibfnamefont{A.}~\bibnamefont{Derevianko}}\ and\ \bibinfo
  {author} {\bibfnamefont{H.}~\bibnamefont{Katori}},\ }%
  \enquote{\bibinfo {title} {{Colloquium: Physics of optical lattice
  clocks}},}\  (\bibinfo {year} {2010}),\
  \Eprint{http://arxiv.org/abs/1011.4622}{arXiv:1011.4622}%
  \bibAnnoteFile{NoStop}{Derevianko2010}%
\bibitem{Katori2003}%
  \BibitemOpen
  \bibfield{author}{%
  \bibinfo {author} {\bibfnamefont{H.}~\bibnamefont{Katori}}, \bibinfo {author}
  {\bibfnamefont{M.}~\bibnamefont{Takamoto}}, \bibinfo {author}
  {\bibfnamefont{V.~G.}\ \bibnamefont{Pal'chikov}},\ and\ \bibinfo {author}
  {\bibfnamefont{V.~D.}\ \bibnamefont{Ovsiannikov}},\ }%
  \bibfield{journal}{%
  \Doi{10.1103/PhysRevLett.91.173005}{\bibinfo {journal} {Phys. Rev. Lett.}}\
  }%
  \textbf{\bibinfo {volume} {91}},\ \bibinfo {pages} {173005} (\bibinfo {year}
  {2003})%
  \bibAnnoteFile{NoStop}{Katori2003}%
\bibitem{Beloy2009}%
  \BibitemOpen
  \bibfield{author}{%
  \bibinfo {author} {\bibfnamefont{K.}~\bibnamefont{Beloy}}, \bibinfo {author}
  {\bibfnamefont{A.}~\bibnamefont{Derevianko}}, \bibinfo {author}
  {\bibfnamefont{V.~A.}\ \bibnamefont{Dzuba}},\ and\ \bibinfo {author}
  {\bibfnamefont{V.~V.}\ \bibnamefont{Flambaum}},\ }%
  \bibfield{journal}{%
  \Doi{10.1103/PhysRevLett.102.120801}{\bibinfo {journal} {Phys. Rev. Lett.}}\
  }%
  \textbf{\bibinfo {volume} {102}},\ \bibinfo {pages} {120801} (\bibinfo {year}
  {2009})%
  \bibAnnoteFile{NoStop}{Beloy2009}%
\bibitem{Rosenbusch2009}%
  \BibitemOpen
  \bibfield{author}{%
  \bibinfo {author} {\bibfnamefont{P.}~\bibnamefont{Rosenbusch}}, \bibinfo
  {author} {\bibfnamefont{S.}~\bibnamefont{Ghezali}}, \bibinfo {author}
  {\bibfnamefont{V.~A.}\ \bibnamefont{Dzuba}}, \bibinfo {author}
  {\bibfnamefont{V.~V.}\ \bibnamefont{Flambaum}}, \bibinfo {author}
  {\bibfnamefont{K.}~\bibnamefont{Beloy}},\ and\ \bibinfo {author}
  {\bibfnamefont{A.}~\bibnamefont{Derevianko}},\ }%
  \bibfield{journal}{%
  \Doi{10.1103/PhysRevA.79.013404}{\bibinfo {journal} {Phys. Rev. A}}\ }%
  \textbf{\bibinfo {volume} {79}},\ \bibinfo {pages} {013404} (\bibinfo {year}
  {2009})%
  \bibAnnoteFile{NoStop}{Rosenbusch2009}%
\bibitem{Derevianko2010a}%
  \BibitemOpen
  \bibfield{author}{%
  \bibinfo {author} {\bibfnamefont{A.}~\bibnamefont{Derevianko}},\ }%
  \bibfield{journal}{%
  \Doi{10.1103/PhysRevLett.105.033002}{\bibinfo {journal} {Phys. Rev. Lett.}}\
  }%
  \textbf{\bibinfo {volume} {105}},\ \bibinfo {pages} {033002} (\bibinfo
  {month} {Jul}\ \bibinfo {year} {2010})%
  \bibAnnoteFile{NoStop}{Derevianko2010a}%
\bibitem{Chicireanu2011}%
  \BibitemOpen
  \bibfield{author}{%
  \bibinfo {author} {\bibfnamefont{R.}~\bibnamefont{Chicireanu}}, \bibinfo
  {author} {\bibfnamefont{K.~D.}\ \bibnamefont{Nelson}}, \bibinfo {author}
  {\bibfnamefont{S.}~\bibnamefont{Olmschenk}}, \bibinfo {author}
  {\bibfnamefont{N.}~\bibnamefont{Lundblad}}, \bibinfo {author}
  {\bibfnamefont{A.}~\bibnamefont{Derevianko}},\ and\ \bibinfo {author}
  {\bibfnamefont{J.~V.}\ \bibnamefont{Porto}},\ }%
  \bibfield{journal}{%
  \Doi{10.1103/PhysRevLett.106.063002}{\bibinfo {journal} {Phys. Rev. Lett.}}\
  }%
  \textbf{\bibinfo {volume} {106}},\ \bibinfo {pages} {063002} (\bibinfo {year}
  {2011})%
  \bibAnnoteFile{NoStop}{Chicireanu2011}%
\bibitem{davidson.105.093001}%
  \BibitemOpen
  \bibfield{author}{%
  \bibinfo {author} {\bibfnamefont{Y.}~\bibnamefont{Sagi}}, \bibinfo {author}
  {\bibfnamefont{I.}~\bibnamefont{Almog}},\ and\ \bibinfo {author}
  {\bibfnamefont{N.}~\bibnamefont{Davidson}},\ }%
  \bibfield{journal}{%
  \Doi{10.1103/PhysRevLett.105.093001}{\bibinfo {journal} {Phys. Rev. Lett.}}\
  }%
  \textbf{\bibinfo {volume} {105}},\ \bibinfo {pages} {093001} (\bibinfo {year}
  {2010})%
  \bibAnnoteFile{NoStop}{davidson.105.093001}%
\bibitem{davidson.105.053201}%
  \BibitemOpen
  \bibfield{author}{%
  \bibinfo {author} {\bibfnamefont{Y.}~\bibnamefont{Sagi}}, \bibinfo {author}
  {\bibfnamefont{I.}~\bibnamefont{Almog}},\ and\ \bibinfo {author}
  {\bibfnamefont{N.}~\bibnamefont{Davidson}},\ }%
  \bibfield{journal}{%
  \Doi{10.1103/PhysRevLett.105.053201}{\bibinfo {journal} {Phys. Rev. Lett.}}\
  }%
  \textbf{\bibinfo {volume} {105}},\ \bibinfo {pages} {053201} (\bibinfo {year}
  {2010})%
  \bibAnnoteFile{NoStop}{davidson.105.053201}%
\bibitem{Lewandowski2002}%
  \BibitemOpen
  \bibfield{author}{%
  \bibinfo {author} {\bibfnamefont{H.~J.}\ \bibnamefont{Lewandowski}}, \bibinfo
  {author} {\bibfnamefont{D.~M.}\ \bibnamefont{Harber}}, \bibinfo {author}
  {\bibfnamefont{D.~L.}\ \bibnamefont{Whitaker}},\ and\ \bibinfo {author}
  {\bibfnamefont{E.~A.}\ \bibnamefont{Cornell}},\ }%
  \bibfield{journal}{%
  \Doi{10.1103/PhysRevLett.88.070403}{\bibinfo {journal} {Phys. Rev. Lett.}}\
  }%
  \textbf{\bibinfo {volume} {88}},\ \bibinfo {pages} {070403} (\bibinfo {year}
  {2002})%
  \bibAnnoteFile{NoStop}{Lewandowski2002}%
\bibitem{Rbsecondary}%
  \BibitemOpen
  \bibfield{journal}{%
  \bibinfo {journal} {CCTF 2004, "Recommendation CCTF-1 2004 concerning
  secondary representations of the second},\ \bibinfo {pages} {38}}%
   (\bibinfo {year} {2004})%
  \bibAnnoteFile{NoStop}{Rbsecondary}%
\bibitem{gibble2009}%
  \BibitemOpen
  \bibfield{author}{%
  \bibinfo {author} {\bibfnamefont{K.}~\bibnamefont{Gibble}},\ }%
  \bibfield{journal}{%
  \Doi{10.1103/PhysRevLett.103.113202}{\bibinfo {journal} {Phys. Rev. Lett.}}\
  }%
  \textbf{\bibinfo {volume} {103}},\ \bibinfo {pages} {113202} (\bibinfo {year}
  {2009})%
  \bibAnnoteFile{NoStop}{gibble2009}%
\bibitem{gibble2010}%
  \BibitemOpen
  \bibfield{author}{%
  \bibinfo {author} {\bibfnamefont{K.}~\bibnamefont{Gibble}},\ }%
  \bibfield{journal}{%
  \Doi{10.1103/Physics.3.55}{\bibinfo {journal} {Physics}}\ }%
  \textbf{\bibinfo {volume} {3}},\ \bibinfo {eid} {55} (\bibinfo {year}
  {2010})%
  \bibAnnoteFile{NoStop}{gibble2010}%
\bibitem{Fuchs2002}%
  \BibitemOpen
  \bibfield{author}{%
  \bibinfo {author} {\bibfnamefont{J.~N.}\ \bibnamefont{Fuchs}}, \bibinfo
  {author} {\bibfnamefont{D.~M.}\ \bibnamefont{Gangardt}},\ and\ \bibinfo
  {author} {\bibfnamefont{F.}~\bibnamefont{Lalo\"e}},\ }%
  \bibfield{journal}{%
  \Doi{10.1103/PhysRevLett.88.230404}{\bibinfo {journal} {Phys. Rev. Lett.}}\
  }%
  \textbf{\bibinfo {volume} {88}},\ \bibinfo {pages} {230404} (\bibinfo {year}
  {2002})%
  \bibAnnoteFile{NoStop}{Fuchs2002}%
\bibitem{kleinebuening2010}%
  \BibitemOpen
  \bibfield{author}{%
  \bibinfo {author} {\bibfnamefont{G.}~\bibnamefont{Kleine~B\"uning}}, \bibinfo
  {author} {\bibfnamefont{J.}~\bibnamefont{Will}}, \bibinfo {author}
  {\bibfnamefont{W.}~\bibnamefont{Ertmer}}, \bibinfo {author}
  {\bibfnamefont{C.}~\bibnamefont{Klempt}},\ and\ \bibinfo {author}
  {\bibfnamefont{J.}~\bibnamefont{Arlt}},\ }%
  \bibfield{journal}{%
  \Doi{10.1007/s00340-010-4078-7}{\bibinfo {journal} {Appl. Phys. B}}\ }%
  \textbf{\bibinfo {volume} {100}},\ \bibinfo {pages} {117} (\bibinfo {year}
  {2010})%
  \bibAnnoteFile{NoStop}{kleinebuening2010}%
\bibitem{Ramirez-Martinez2010}%
  \BibitemOpen
  \bibfield{author}{%
  \bibinfo {author} {\bibfnamefont{F.}~\bibnamefont{Ramirez-Martinez}},
  \bibinfo {author} {\bibfnamefont{M.}~\bibnamefont{Lours}}, \bibinfo {author}
  {\bibfnamefont{P.}~\bibnamefont{Rosenbusch}}, \bibinfo {author}
  {\bibfnamefont{F.}~\bibnamefont{Reinhard}},\ and\ \bibinfo {author}
  {\bibfnamefont{J.}~\bibnamefont{Reichel}},\ }%
  \bibfield{journal}{%
  \Doi{10.1109/TUFFC.2010.1383}{\bibinfo {journal} {IEEE Trans. Ultrason.
  Ferroelectr. Freq. Control}}\ }%
  \textbf{\bibinfo {volume} {57}},\ \bibinfo {pages} {88 } (\bibinfo {year}
  {2010})%
  \bibAnnoteFile{NoStop}{Ramirez-Martinez2010}%
\bibitem{Reinaudi2007}%
  \BibitemOpen
  \bibfield{author}{%
  \bibinfo {author} {\bibfnamefont{G.}~\bibnamefont{Reinaudi}}, \bibinfo
  {author} {\bibfnamefont{T.}~\bibnamefont{Lahaye}}, \bibinfo {author}
  {\bibfnamefont{Z.}~\bibnamefont{Wang}},\ and\ \bibinfo {author}
  {\bibfnamefont{D.}~\bibnamefont{Gu\'{e}ry-Odelin}},\ }%
  \bibfield{journal}{%
  \Doi{10.1364/OL.32.003143}{\bibinfo {journal} {Opt. Lett.}}\ }%
  \textbf{\bibinfo {volume} {32}},\ \bibinfo {pages} {3143} (\bibinfo {year}
  {2007})%
  \bibAnnoteFile{NoStop}{Reinaudi2007}%
\bibitem{Guena2010}%
  \BibitemOpen
  \bibfield{author}{%
  \bibinfo {author} {\bibfnamefont{J.}~\bibnamefont{Guena}}, \bibinfo {author}
  {\bibfnamefont{P.}~\bibnamefont{Rosenbusch}}, \bibinfo {author}
  {\bibfnamefont{P.}~\bibnamefont{Laurent}}, \bibinfo {author}
  {\bibfnamefont{M.}~\bibnamefont{Abgrall}}, \bibinfo {author}
  {\bibfnamefont{D.}~\bibnamefont{Rovera}}, \bibinfo {author}
  {\bibfnamefont{G.}~\bibnamefont{Santarelli}}, \bibinfo {author}
  {\bibfnamefont{M.}~\bibnamefont{Tobar}}, \bibinfo {author}
  {\bibfnamefont{S.}~\bibnamefont{Bize}},\ and\ \bibinfo {author}
  {\bibfnamefont{A.}~\bibnamefont{Clairon}},\ }%
  \bibfield{journal}{%
  \Doi{10.1109/TUFFC.2010.1461}{\bibinfo {journal} {IEEE Trans. Ultrason.
  Ferroelectr. Freq. Control}}\ }%
  \textbf{\bibinfo {volume} {57}},\ \bibinfo {pages} {647 } (\bibinfo {year}
  {2010})%
  \bibAnnoteFile{NoStop}{Guena2010}%
\bibitem{Note1}%
  \BibitemOpen
  \bibinfo {note} {We use a$_{\delimiter "3222378 \delimiter "3222378 }\protect
  \tmspace -\thinmuskip {.1667em}=\protect \tmspace -\thinmuskip
  {.1667em}94.55~a_\protect \text {B}$, a$_{\delimiter "3223379 \delimiter
  "3223379 }\protect \tmspace -\thinmuskip {.1667em}=\protect \tmspace
  -\thinmuskip {.1667em}100.76~a_\protect \text {B}$ from a coupled channel
  analysis by E. Tiemann.}%
  \bibAnnoteFile{Stop}{Note1}%
\bibitem{Note2}%
  \BibitemOpen
  \bibinfo {note} {A.~Derevianko, Priv. Comm. (see also \cite
  {Derevianko2010a})}%
  \bibAnnoteFile{NoStop}{Note2}%
\bibitem{Howe1981}%
  \BibitemOpen
  \bibfield{author}{%
  \bibinfo {author} {\bibfnamefont{D.}~\bibnamefont{Howe}}, \bibinfo {author}
  {\bibfnamefont{D.}~\bibnamefont{Allan}},\ and\ \bibinfo {author}
  {\bibfnamefont{J.}~\bibnamefont{Barnes}},\ }%
  \bibfield{journal}{%
  \bibinfo {journal} {Proc. 35th Annu. Freq. Contr. Symp., Philadelphia, PA}}%
   (\bibinfo {year} {1981})%
  \bibAnnoteFile{NoStop}{Howe1981}%
\end{thebibliography}%
\end{document}